\documentclass[12pt,a4paper]{article}
\usepackage{amssymb}
\usepackage[unicode]{hyperref}
\begin{document}
\def\emline#1#2#3#4#5#6{%
       \put(#1,#2){\special{em:moveto}}%
       \put(#4,#5){\special{em:lineto}}}
\def\newpic#1{}
\title{The origin of the energy-momentum conservation law}
\author{Andrew E Chubykalo${}^\dagger$, Augusto Espinoza${}^\dagger$,
and B P Kosyakov${}^{\ddagger \S}$}
\maketitle
\begin{center}
{\small${}^\dagger$ 
Escuela de F{\'i}sica, 
Universidad Aut\'onoma de Zacatecas, 
Apartado Postal C-580 Zacatecas 98068, Zacatecas,  
Mexico\\
${}^\ddagger$ Russian Federal Nuclear Center, 
Sarov, 607189 Nizhni\u{\i} Novgorod Region, Russia \\
${}^\S$ Moscow Institute of Physics {\&} Technology, Dolgoprudni\u{\i}, 141700 Moscow Region, 
Russia\\
{\bf E-mail:} 
${\rm achubykalo@yahoo.com.mx}$, 
${\rm drespinozag@yahoo.com.mx}$, 
${\rm kosyakov.boris@gmail.com}$} 
\end{center}

\begin{abstract}
{
The interplay between 
the action--reaction principle and the energy-momentum conservation law
is revealed by the examples of the Maxwell--Lorentz and Yang--Mills--Wong 
theories, and general relativity.
These two statements are shown to be equivalent in the sense that both hold or 
fail together.
Their mutual agreement is demonstrated most clearly in the self-interaction 
problem by taking account of the rearrangement of degrees of freedom 
appearing in the action of the Maxwell--Lorentz and Yang--Mills--Wong theories.
The failure of energy-momentum conservation in general relativity is attributed 
to the fact that this theory allows solutions having nontrivial topologies.
The total energy and momentum of a system with nontrivial topological content
prove to be ambiguous, coordinatization-dependent quantities.
For example, the energy of a Schwarzschild black hole may take any positive 
value greater than, or equal to, the mass of the  body whose  collapse is 
responsible for arising this black hole.
We draw the analogy to the paradoxial Banach--Tarski theorem; the measure 
becomes a poorly defined concept if initial three-dimensional bounded sets are 
rearranged in topologically nontrivial ways through the action of free 
non-Abelian isometry groups.
}
\end{abstract}

\noindent
Keywords: action--reaction, translation invariance, energy and momentum conservation,
rearrangement of initial degrees of freedom

\section{Introduction}                            
\label
{Introduction}
By summing the basic advances in physics of the 19th century, Max Planck 
placed strong emphasis on the action--reaction principle as the rationale 
of momentum conservation \cite{Planck}.
On the other hand, following Noether's first theorem \cite{Noether}, we recognize 
that any dynamical system exhibits momentum conservation if the action of 
this system is invariant under space translations, in other words, the 
momentum conservation law stems from homogeneity of space.

In nonrelativistic mechanics, Newton's third law is consistent with the 
requirement of translation invariance. 
Indeed, the forces exerted on particles in an isolated two-particle system are on 
the same line, equal, and oppositely directed when the potential energy 
assumes the form $U({\bf z}_1-{\bf z}_2)$, where ${\bf z}_1$ and ${\bf z}_2$ 
are coordinates of these particles.
However, this law is no longer valid in relativistic mechanics where the 
influence of one particle on another propagates at a finite speed, and the 
response arises with some retardation.
Furthermore, energy and momentum are fused into energy-momentum whose 
conservation is attributed to homogeneity of Minkowski spacetime.
So the Planck's insight into the reason for momentum conservation is gradually
fading from the collective consciousness of theoretical physics.

Meanwhile there is one exceptional case, namely contact interactions, in which 
one particle
acts on another and experiences back reaction at the same point, as
exemplified by collisions and decays of pointlike particles.
This form of relativistic interactions respects both Newton's third law and 
energy-momentum conservation, suggesting to consider 
the action--reaction principle in a broader sense and extend it to cover
local interactions in classical field theories.
The most familiar example can be found in the Maxwell--Lorentz electrodynamics 
in which the role of the electric charge $e$ is twofold: $e$ acts as both coupling 
between the point particle carrying this charge and 
electromagnetic field  
and the strength of the delta-function source of electromagnetic field.

To gain a clearer view of whether the action--reaction principle has a 
direct bearing on energy--momentum conservation, one should invoke the 
self-interaction problem.
This issue is studied in Sects.~2 and 3 by the examples 
of Maxwell--Lorentz  electrodynamics and Yang--Mills--Wong theory.

Turning to general relativity in Sect.~4, we conclude that both action--reaction 
principle and energy--momentum conservation cease to be true.
The absence of energy-momentum conservation from this theory is due to the fact
that the equation of gravitational field allows solutions which represent
spacetime manifolds with nontrivial topology.
Energy and momentum may thus become poorly defined concepts in general relativity.
It transpires that the total energy of a Schwarzschild black hole may 
take any positive value greater than, or equal to, the mass of the collapsed 
body in different coordinatizations.
The situation closely resembles that in the paradoxial Banach--Tarski theorem.
We sketch the broad outline of this theorem and its potential relevance to the 
problem of poorly defined measure for total energy and momentum in Sects.~4 and 5.
The rearrangement of degrees of freedom appearing in the action and its
role in facilitating the integral
quantities to become well-defined is discussed in Sect.~5.

We follow the notation used in \cite{k07}.
In Sects.~2, 3, and 5, in which our attention is restricted essentially to the 
picture in Minkowski spacetime, we adopt the mainly negative signature $(+---)$ 
convenient to the description of world lines. 
In Sect.~4, we proceed from the idea of pseudo-Riemannian spacetime, and
use the mainly positive signature $(-+++)$, which is particularly adapted to the
description of 3-dimensional surfaces.
We put the speed of light equal to unity throughout. 

\section{The Maxwell--Lorentz electrodynamics}
\label
{Maxwell}
The action  
\begin{equation}
{S}
=
-\int d^4x
\left(\frac{1}{16\pi}\,F_{\mu\nu} F^{\mu\nu}+j^\mu A_\mu\right)
-m_0\int d\tau\,\sqrt{{\dot z}^\mu\,{\dot z}_\mu}
\label
{action-Mxw-Lorntz}
\end{equation}                        
encodes the dynamics of a single charged particle interacting 
with electromagnetic field.
Here, 
\begin{equation}
j^\mu(x)
=
e\int^\infty_{-\infty} d\tau\,{\dot z}^\mu(\tau)\,\delta^{4}\!\left[x-z(\tau)\right]
\label
{j-mu-single}
\end{equation}                        
is the current density produced by the 
particle moving along a smooth timelike world line $z^\mu(\tau)$ and carrying the 
charge $e$, and $m_0$ is the mechanical mass of this {bare} particle.

A closed system of this kind enjoys the property of translational invariance which 
affords energy-momentum conservation through the famous Noether argument.  

The comparison of the source, Eq.~(\ref{j-mu-single}), in the field equation 
\begin{equation}
{\cal E}_\mu=\partial^\nu F_{\mu\nu} 
+4\pi j_\mu =0\, 
\label
{Eulerian-em-II}
\end{equation}                        
with the Lorentz force in the equation of motion for this charged particle 
\begin{equation}
\varepsilon^\lambda=m_0 {\ddot z}^\lambda-
e {\dot z}_\mu F^{\lambda\mu}=0\,,
\label
{Eulerian-part}
\end{equation}                        
where the dot stands for the derivative with respect to the proper time $s$ of
the particle, shows that both are scaled by the same parameter $e$.
This fact is consistent with the action--reaction principle: 
$e$ measures both variation of the particle state for a given field state and 
variation of the field state for a given particle state. 

Does this statement bear on energy-momentum conservation?   
To answer this question, we turn to the {\it self-interaction} problem.
Naively, this problem is about interfacing the bare particle and electromagnetic 
field on the world line, which will hopefully reveal local energy-momentum 
balance of  this contact interaction.
We are therefore to address a simultaneous solution of 
equations (\ref{Eulerian-em-II}) and (\ref{Eulerian-part}). 
To see this, consider the Noether identity 
\begin{equation}
\partial_\mu T^{\lambda\mu}
=
{1\over 4\pi}\,{\cal E}_\mu F^{\lambda\mu}
+
\int^\infty_{-\infty}ds\,\varepsilon^\lambda(z)\,\delta^4\left[x-z(s)\right],
\label
{Noether-1-id-em}
\end{equation}                         
where $T^{\mu\nu}$ is the total metric stress-energy tensor of this system, 
\begin{equation}
T^{\mu\nu}
=
\frac{2}{\sqrt{-g}}\frac{\delta S}{\delta g^{\mu\nu}}
=
\Theta^{\mu\nu}+t^{\mu\nu}\,,
\label
{T=Theta+t}
\end{equation}                         
\begin{equation}
\Theta^{\mu\nu}={1\over 4\pi}\left(F^{\mu\alpha}F_\alpha^{\hskip 1.5mm
\nu} +{1\over 4}\,\eta^{\mu\nu}F_{\alpha\beta}F^{\alpha\beta}\right),
\label
{Theta-mu-nu}
\end{equation}                           
\begin{equation}
t^{\mu\nu}= m_0\int^\infty_{-\infty}\! ds\,{\dot z}^\mu(s)\,{\dot z}^\nu(s)\,
\delta^4\left[x-z(s)\right],
\label
{t-mu-nu}
\end{equation}                         
and  ${\cal E}_\mu$ and $\varepsilon^\lambda$ are, 
respectively, the left-hand sides of equations  (\ref{Eulerian-em-II})
and (\ref{Eulerian-part}).
It follows from (\ref{Noether-1-id-em}) that ${\cal E}_\mu=0$ and $\varepsilon^\lambda=0$ 
imply $\partial_\mu T^{\lambda\mu}=0$, that is, the equation of motion for a 
bare particle (\ref{Eulerian-part}), in which an appropriate solution to the 
field equation (\ref{Eulerian-em-II}) has been used, is equivalent 
to the local conservation law for the total stress-energy tensor. 

Imposing the retarded boundary condition, we obtain a solution to 
equation (\ref{Eulerian-em-II}) in the Li\'enard--Wiechert form,  
\begin{equation}
F^{\mu\nu}_{\rm ret}=\frac{e}{\rho^2}\left({R}^\mu{V}^\nu-{R}^\nu{V}^\mu\right),
\label
{F-mu-nu-LW-III}
\end{equation}                          
\begin{equation}
V^\mu=\left[1-(R\cdot {\ddot z})\right]\frac{{\dot z}^\mu}{\rho}+{\ddot z}^\mu\,,
\label
{V-}
\end{equation}                       
$R^\mu=x^\mu-{z}^\mu(s_{\rm ret})$ is a lightlike vector drawn from a point 
${z}^\mu(s_{\rm ret})$ on the world line, where the signal was emitted, to the 
point $x^\mu$, where the signal was received, and $\rho=R\cdot{\dot z}$ is the 
spatial distance between  $x^\mu$ and ${z}^\mu(s_{\rm ret})$ in the 
instantaneously comoving Lorentz frame in which the charge is at rest at the 
retarded instant $s_{\rm ret}$.
The field (\ref{F-mu-nu-LW-III})--(\ref{V-}) is singular on the world line.
Substituting it into (\ref{Eulerian-part}) results in a divergent expression. 
This divergence is a manifestation of infinite {self-interaction}: the charged 
bare particle experiences its own electromagnetic field which is infinite at the 
point of origin. 

A possible cure for this difficulty is to regularize the Li\'enard--Wiechert 
field $F^{\mu\nu}_{\rm ret}$ in a small vicinity of the world line. 
Take, for example, the field as a function of two variables 
$F^{\mu\nu}_{\rm ret}(x;z(s_{\rm ret}))$ and continue it analytically from null intervals between 
the observation points $x^\mu$ and the retarded points ${z}^\mu(s_{\rm ret})$ 
to timelike intervals that result from assigning $x^\mu=z^\mu(s_{\rm ret}+\epsilon)$ and
keeping the second variable $z^\mu(s_{\rm ret})$ fixed \cite{Barut}. 
A crucial step in removing the regularization is to change $m_0$ to a function 
of regularization, $m_0(\epsilon)$, add it to the divergent term ${e^2}/{2\epsilon}$, 
and assume that 
\begin{equation}
m=\lim_{\epsilon\to 0}\left[m_0(\epsilon)+ \frac{e^2}{2\epsilon}\right]
\label
{m-ren}
\end{equation}                        
is finite and positive.
Then the divergence disappears, and we arrive at the Lorentz--Dirac equation \cite{Dirac}
\begin{equation} 
{\Lambda}^\mu=m {\ddot z}^{\hskip0.3mm\mu}-{2\over 3}\,e^2\left({\stackrel{\ldots}z}^{\hskip0.3mm\mu}
+{\dot z}^\mu {\ddot z}^2\right)-f^{\mu}_{\rm ext}=0\,,
\label
{LD}
\end{equation}                       
where $f^{\mu}_{\rm ext}=e{\dot z}_\nu F^{\mu\nu}_{\rm ext}$ is an external 
four-force, with $F^{\mu\nu}_{\rm ext}$ being a free electromagnetic field.

Is it possible to regard (\ref{LD}) as the desired equation of local energy-momentum 
balance?
Based on the wide-spread belief that the Abraham term
\begin{equation}
\Gamma^\mu= 
{2\over 3}\,e^2\left({\stackrel{\ldots}z}^{\hskip0.3mm\mu}+
{\dot z}^\mu{\ddot z}^2\right)
\label
{Abraham-term-def}
\end{equation}                                           
is the radiation reaction {\it four-force}, one would give a negative answer to 
this question.
This is because the radiating particle feels a recoil equal to the negative of 
the Larmor emission rate 
\begin{equation} 
{\dot{\cal P}}^\mu=-{2\over 3}\,e^2\,{\dot z}^\mu{\ddot z}^2\,.
\label
{neg-Larmor}
\end{equation}
However, $-{\dot{\cal P}}^\mu$ cannot be considered as a four-force because it
is not orthogonal to ${\dot z}^\mu$.
On the other hand, $\Gamma^\mu$ is orthogonal to ${\dot z}^\mu$, but it differs
from the anticipated recoil by the so-called Schott term
${2\over 3}\,e^2\,{\stackrel{\ldots}z}^{\hskip0.3mm\mu}$.
Although the energy stored in the Schott term can be attributed to a 
reversible form of emission and absorption of field energy \cite{Dirac}, 
its actual role appears mysterious.

Furthermore, the general solution to equation (\ref{LD}) with $f^{\mu}_{\rm ext}=0$ 
is
\begin{equation} 
{\dot z}^\mu (s)=e_0^\mu\cosh(\alpha_0+w_0\tau_0\, e^{s/\tau_0})
+
e_1^\mu\sinh(\alpha_0+w_0\tau_0\,e^{s/\tau_0})\,,
\label
{self-acceleration}
\end{equation}                         
where $e_0^\mu$ and $e_1^\mu$ are constant vectors such that $e_0\cdot e_1=0$,  
$e_0^2=-e_1^2=1$, $\tau_0=2e^2/3m$, $\alpha_0$ and $w_0$ are arbitrary constants.
The solution (\ref{self-acceleration}) is an embarrassing feature of the 
Lorentz--Dirac equation:
a free charged particle moving along this world line continually accelerates,
\begin{equation}
{\ddot z}^2(s)=-w_0^2\exp\left(2s/\tau_0\right),
\label
{a^2-runaway}
\end{equation}
and continually radiates.
This {\it self-acceleration} seems {\it contrary to the energy-momentum conservation 
law} even though this law is assured by {\it translational invariance} of the action.

These paradoxial results\footnote{For other paradoxes related to self-interaction 
in the Maxwell--Lorentz theory see, e.~g., \cite{Rohrlich}.} 
signal that self-interaction is a subtle issue whose treatment 
requires further refinements of the conceptual basis.
A plausible assumption is that the extremization of the action, subject to the 
retarded condition, may result in {unstable} modes, which culminates in  
{\it rearranging} the initial degrees of freedom \cite{k07}.
The action (\ref{action-Mxw-Lorntz}) is expressed in terms of mechanical 
variables $z^\mu(\tau)$ describing world lines of a bare charged particle and 
the electromagnetic vector potential $A^\mu(x)$.
The rearrangement of these degrees of freedom yields new 
dynamically independent entities, a {\it dressed} charged {particle} and {\it 
radiation}.

We begin with the local conservation law for the total stress-energy tensor   
\begin{equation}
\partial_\lambda T^{\lambda\mu}=0\,.
\label
{Total-conserv-law}
\end{equation}                         
Recall that taking the local conservation law (\ref{Total-conserv-law}), as the 
starting point in the self-energy analysis, is as good as that of 
simultaneous solution of dynamical equations  (\ref{Eulerian-em-II}) and (\ref{Eulerian-part}).
Substituting the general solution of the field equation 
(\ref{Eulerian-em-II}), $F^{\mu\nu}=F^{\mu\nu}_{\rm ret}+F^{\mu\nu}_{\rm ext}$,
into (\ref{Theta-mu-nu}) gives
\begin{equation}
\Theta^{\mu\nu}= -{e^2\over 4\pi\rho^4}\left[V^2 R^\mu R^\nu
-
\left(R^\mu V^\nu+R^\nu V^\mu\right)
+
{1\over 2}\,\eta^{\mu\nu}\right]
+
\Theta_{\rm mix}^{\mu\nu}\,, 
\label
{Theta-via-LW}
\end{equation}                         
where the first term results from the self-field (\ref{F-mu-nu-LW-III})--(\ref{V-}),
and the second term contains mixed contributions of the self-field and free field.
The first term splits into two parts $\Theta_{\rm bound}^{\mu\nu}+\Theta_{\rm rad}^{\mu\nu}$,
where
\begin{equation}
\Theta^{\mu\nu}_{\rm bound}=-{e^2\over 4\pi\rho^4}\left[\frac{R^\mu R^\nu}{\rho^2}\left(1
-2R\cdot{\ddot z}\right)-\left(R^\mu V^\nu+R^\nu V^\mu\right) 
+
{1\over 2}\,\eta^{\mu\nu}\right],
\label
{Theta-I}
\end{equation}                         
\begin{equation}
\Theta^{\mu\nu}_{\rm rad}
=
-{e^2\over 4\pi\rho^4}\left[{\ddot z}^2
+
\frac{1}{\rho^2}\left({\ddot z}\cdot R\right)^2\right]R^\mu R^\nu\,.
\label
{Theta-II}
\end{equation}                           
The following local conservation laws hold off the world line \cite{Teitelboim}:
\begin{equation}
\partial_\mu\Theta^{\mu\nu}_{\rm bound}=0\,,
\quad
\partial_\mu \Theta_{\rm rad}^{\mu\nu}=0\,,
\quad
\partial_\mu \Theta_{\rm mix}^{\mu\nu}=0\,.
\label
{cons-Theta-I-II}
\end{equation}                                       
A natural interpretation of (\ref{cons-Theta-I-II}) is that 
$\Theta^{\mu\nu}_{\rm bound}$,  $\Theta_{\rm rad}^{\mu\nu}$, and $\Theta^{\mu\nu}_{\rm mix}$
are {\it dynamically independent}  outside the world line  \cite{Teitelboim}.
There is no other decomposition of $\Theta^{\mu\nu}$ into parts which may 
be recognized as dynamically independent.

Since $\Theta_{\rm rad}^{\mu\nu}$ and $\Theta^{\mu\nu}_{\rm mix}$ behave like 
$\rho^{-2}$ near the world line, they are
integrable over a three-dimensional spacelike surface $\Sigma$, and, 
in view of (\ref{cons-Theta-I-II}), the surface of integration may 
be deformed from $\Sigma$ to more geometrically motivated surfaces.
It is convenient to substitute $\Sigma$ by a tube $T_\epsilon$ of infinitesimal radius 
$\epsilon$ enclosing the world line to obtain
\begin{equation}
{\cal P}^{\mu}=
\int_{\Sigma}d\sigma_\lambda\,\Theta^{\lambda\mu}_{\rm rad}
=
\lim_{\epsilon\to 0}\int_{T_\epsilon}d\sigma_\lambda\,\Theta^{\lambda\mu}_{\rm rad}
=
-{2\over 3}\,e^2\int^{{s}}_{-\infty}d\tau\,{\ddot z}^2(\tau) {\dot z}^\mu(\tau)\,
\label
{P-rad-def}
\end{equation}     
and
\begin{equation}
{\wp}^{\mu}
=
\int_{\Sigma}d\sigma_\lambda\,\Theta^{\lambda\mu}_{\rm mix}
=
\lim_{\epsilon\to 0}\int_{T_\epsilon}d\sigma_\lambda\,\Theta^{\lambda\mu}_{\rm mix}
=
-e\int^{{s}}_{-\infty}d\tau\,F_{\rm ext}^{\mu\nu}(z)\,{\dot z}_\nu(\tau)\,.
\label
{P-mix}
\end{equation}                        

${\cal P}^{\mu}$ represents the four-momentum radiated by the charge $e$ during the whole past history prior to the instant $s$.
Indeed: (i) $\Theta^{\mu\nu}_{\rm rad}$ is a dynamically independent part of 
$\Theta^{\mu\nu}$; (ii) $\Theta^{\mu\nu}_{\rm rad}$ moves away from the charged 
particle with the speed of light, more precisely, $\Theta^{\mu\nu}_{\rm rad}$ 
propagates along the future light cone $C_+$ drawn from the emission point, 
$\Theta^{\mu\nu}_{\rm rad}R_\nu=0$; 
(iii) the flux of $\Theta^{\mu\nu}_{\rm rad}$ goes as $\rho^{-2}$ implying that the 
same amount of  energy-momentum flows through
spheres of different radii. 
Differentiating  (\ref{P-rad-def}) with respect to $s$ gives the Larmor 
four-momentum emitted by the accelerated charge per 
unit proper time, Eq.~(\ref{neg-Larmor}).

As for ${\wp}^{\mu}$, it is the four-momentum extracted from the {free} field $F_{\rm ext}^{\mu\nu}(x)$ during the whole past history up to the 
instant $s$. 

By contrast, $\Theta_{\rm bound}^{\mu\nu}$ contains singularities 
$\rho^{-3}$ and $\rho^{-4}$ which are not integrable.
Hence, an appropriate regularization is necessary. 
For example, employing a Lorentz-invariant cutoff prescription \cite{k07}, one 
finds  
\begin{equation}
P^\mu_{\rm bound}
=
{\rm Reg}_\epsilon\int_{\Sigma}d\sigma_\lambda\,\Theta_{\rm bound}^{\lambda\mu}
=
{e^2\over 2\epsilon}\,{\dot z}^{\mu}-{2\over 3}\,e^2 {\ddot z}^{\mu}\,,
\label
{P-bound}
\end{equation}                        
where $\epsilon$ is the cutoff parameter which must go to zero in the end of
calculations.
Since the flux of $\Theta_{\rm bound}^{\mu\nu}$ through $C_+$ is nonzero,
$\Theta^{\mu\nu}_{\rm bound}R_\nu\ne 0$, $\Theta_{\rm bound}^{\mu\nu}$ propagates slower than 
light.
Unlike $\Theta^{\mu\nu}_{\rm rad}$, which detaches from the source,  $\Theta_{\rm bound}^{\mu\nu}$ 
remains bound to the source \cite{Teitelboim}. 
In other words, the source carries the four-momentum $P^\mu_{\rm bound}$ 
along with its motion.

From (\ref{P-bound}) follows that the measure $d\sigma_\lambda\Theta_{\rm bound}^{\lambda\mu}$
is ill-defined.
However, observing that
\begin{equation}
p^\mu_{0}
=
\int_{\Sigma}d\sigma_\lambda\,t^{\lambda\mu}
=
{m_0}{\dot z}^{\mu}\,,
\label
{p-0}
\end{equation}                        
one may render $m_0$ a singular function of $\epsilon$, $m_0(\epsilon)$, 
add Eqs.~(\ref{P-bound}) and (\ref{p-0}) up, and carry out the renormalization 
of mass, Eq.~(\ref{m-ren}), to complete the definition of the measure 
${\rm Reg}_\epsilon\,d\sigma_\lambda\left(
\Theta_{\rm bound}^{\lambda\mu}+t^{\lambda\mu}\right)$  in the limit $\epsilon\to 0$, 
and eventually arrive at 
\begin{equation}
p^\mu
=
\lim_{\epsilon\to 0}{\rm Reg}_\epsilon\int_{\Sigma}d\sigma_\lambda\left(
\Theta_{\rm bound}^{\lambda\mu}+t^{\lambda\mu}\right)
=
m {\dot z}^{\mu}-{2\over 3}\,e^2 {\ddot z}^{\mu}\,.
\label
{p-dressed}
\end{equation}                        
This four-momentum, originally deduced in  \cite{Teitelboim}, is to be 
attributed to the {\it dressed} particle. 

We now  integrate (\ref{Total-conserv-law}) over a 
domain of spacetime bounded by two spacelike surfaces ${\Sigma}'$ and 
${\Sigma}''$, separated by a small timelike interval, with both normals 
directed towards the future, and a tube ${T}_{R}$ of large radius ${R}$.
Applying the Gau{ss}--Ostrogradski\v{\i} theorem, we obtain\footnote{We assume
that $F_{\rm ext}^{\lambda\mu}(x)$ disappears at spatial infinity.
Therefore, the only term contributing to the integral over $T_R$ is
$\Theta^{\mu\nu}_{\rm rad}$.
Taking into account the second equation of  (\ref{cons-Theta-I-II}), the integral of 
$\Theta^{\mu\nu}_{\rm rad}$ over $T_R$ can be converted into the integral  
over $T_\epsilon$, so that the upshot is given by Eq.~(\ref{P-rad-def}).}:
\[
\left(\int_{{\Sigma}''}-\int_{{\Sigma}'}+\int_{{T}_{R}}\right)\,d\sigma_\mu
\left(\Theta^{\lambda\mu}+t^{\lambda\mu}\right)
\]
\[
=
\left\{\lim_{\epsilon\to 0}\left[m_0(\epsilon)+{e^2\over 2\epsilon}\right]{\dot z}^{\lambda}
-
{2\over 3}\,e^2 {\ddot z}^{\lambda}\right\}\Delta s
-
\int_{{s}}^{{s+\Delta s}}d\tau\left[{2\over 3}\,e^2\,{\ddot z}^2(\tau) {\dot z}^\lambda(\tau)
+
eF_{\rm ext}^{\lambda\mu}(z)\,{\dot z}_\mu(\tau)\right]=0\,,
\]
or, in a concise form  \cite{k07},
\begin{equation}
\Delta p^\lambda
+
\Delta {\cal P}^\lambda
+
\Delta{\wp}^\lambda
=0\,.
\label
{balance-LD}
\end{equation}                        
Evidently (\ref{balance-LD}) is identical to the Lorentz--Dirac equation (\ref{LD}).

On the other hand,  (\ref{balance-LD}) is the desired {\it energy-momentum 
balance}: the four-momentum 
$\Delta{\wp}^{\lambda}=-eF_{\rm ext}^{\lambda\mu}\,{\dot z}_\mu\Delta s$ 
which is extracted from the external field $F_{\rm ext}^{\lambda\mu}$ during 
the period of time 
$\Delta s$ is distributed between the 
four-momentum of the dressed particle $\Delta p^{\lambda}$ and the four-momentum 
carried away by radiation ${\Delta{{\cal P}}}^\lambda$.

Of particular interest is the case $F_{\rm ext}^{\lambda\mu}=0$,
\begin{equation}
\Delta p^\lambda
=
-
\Delta {\cal P}^\lambda\,.
\label
{self-int-em-rearr}
\end{equation}                        
It immediately follows that the rate of change of the energy-momentum of a 
dressed particle, ${\dot p}^\lambda$, is equal to the negative of the Larmor
emission rate, $-{\dot{\cal P}}^\lambda$.
Here, two remarks are in order.
First, $-\Delta {\cal P}^\lambda$ is a mere four-momentum\footnote{A possible 
interpretation of Eq.~(\ref{self-int-em-rearr}) is that the dressed particle 
experiences a jet thrust in response to emitting the electromagnetic field 
momentum, a kind of apparent force applied to the same point in which the 
emission occurs.} 
(rather than four-force), and hence the fact that ${\dot{\cal P}}^\lambda$ is 
not orthogonal to ${\dot z}^\lambda$ presents no special problem.
Second, the energy of a dressed particle is {\it indefinite}\footnote{The fact 
that $p^0$ is not positive definite is scarcely surprising.
Recall that $p^\mu$ is the sum of two vectors $p^\mu=m_0 {\dot z}^\mu+P^\mu_{\rm bound}$.
The bound four-momentum $P^\mu_{\rm bound}$ is a timelike future-directed vector, 
while the four-momentum of a bare particle $m_0 {\dot z}^\mu$ is a timelike 
past-directed vector because $m_0(\epsilon)<0$ for small $\epsilon$,
as (\ref{m-ren}) suggests.
Assuming that $m_0 {\dot z}^\mu+P^\mu_{\rm bound}$ is a timelike vector, one 
recognizes that the time component of this vector can have any sign.},
\begin{equation}
p^{0}=m\gamma \left(1-\tau_0\gamma ^{3}{\bf a}\cdot {\bf v}\right),
\label
{p-0-em-}
\end{equation}
where  $\gamma$ is the Lorentz factor  $\gamma=(1-{\bf v}^2)^{-1/2}$.
Therefore, increasing $|{\bf v}|$ need not be accomplished by  
increasing $p^0$.
For instance, one may readily check that the energy of a dressed particle executing a 
self-accelerated motion (\ref{self-acceleration}) steadily decreases, which 
exactly compensates the increase in energy of the electromagnetic field 
emitted\footnote{
The solution (\ref{self-acceleration}) is usually thought of as a pathological 
trait of the Lorentz--Dirac equation (\ref{LD}) for two main reasons: (i) this 
solution seems incompatible with energy conservation, and (ii) there is no 
experimental evidence for self-accelerated motions in the Nature.
Both accusations are unjust.
The fact that energy-momentum is conserved in this motion has just now been 
established.
As to the manifestation of this phenomenon, the universe as a whole exhibiting accelerated expansion provides an 
excellent potential example of a free entity (brane?) which executes exponentially 
accelerated motion with the characteristic time equal to the inverse of current Hubble 
scale and emits gravitational radiation \cite{k05}. 
Why is the self-accelerated motion of charged particles never observed?
It follows from (\ref{p-dressed}) that 
\begin{equation}
{p^2}=m^2\left({1+\tau_0^2\,a^2}\right).
\label
{M-m}
\end{equation}                          
If $\tau_0^2\,a^2<-1$, the dressed  charged particle turns to a tachyonic 
state $p^2<0$.
Let the particle be moving in the self-accelerated regime (\ref{self-acceleration}).
Then, after a lapse of time $\Delta t=-{\tau_0}\,\log\,\tau_0|{w_0}|$, the 
critical acceleration $\vert{a}^2\vert=\tau_0^{-2}$ is exceeded, 
and the four-momentum of this dressed particle becomes spacelike.
The period of time $\Delta t$ over which a self-accelerated electron possesses 
timelike four-momenta is estimated at $\tau_0\sim 10^{-23}$ s 
for electrons, and still shorter for more massive charged elementary particles.
All primordial self-accelerated particles with such $\tau_0$'s have long 
been in the tachyonic state.
However, we have not slightest notion of how tachyons can be experimentally 
recorded \cite{k05}.
Noteworthy also is that non-Galilean and Galilean regimes of motion are never 
interconvertible: the history of a particular dressed charged particle is decided 
by the asymptotic condition in the limit $s\to\infty$. 
It is then conceivable that the Galilean form of evolution, corresponding 
to $w_0=0$ in Eq.~(\ref{self-acceleration}), may well be assigned to all 
dressed charged particles.  
}.

\section{The Yang--Mills--Wong theory}
\label
{YMW}
The Yang--Mills--Wong theory describes the classical interaction of particles 
carrying non-Abelian charges with the corresponding Yang--Mills field \cite{Wong}.
A system of $K$ such particles 
(thereafter called quarks) interacting with the SU$({\cal N})$ 
Yang--Mills field is governed by the action \cite{Balachandran}
\[
S=-\sum_{I=1}^K\,\int d\tau_I\left\{m^I_0\,\sqrt{{\dot z}_I\cdot{\dot z}_I}
+\sum_{a=1}^{{\cal N}^2-1}\sum_{i,j=1}^{{\cal N}}\, q_I^a\,\eta_{Ii}^{\ast}\left[\delta^i_{\hskip0.5mm j}\frac{d}{d\tau_I}
+
{\dot z}_I^\mu\!\left( A^a_{~\mu} T_a\right)^{i}_{\hskip0.5mm j}\right]
\eta_{I}^{\hskip0.3mm j}\right\}
\]
\begin{equation}
-\frac{1}{16\pi}\int d^4x
\,G_a^{\mu\nu}G^a_{\mu\nu}\,,
\label
{Balachandr-action}
\end{equation}                                           
where $T_a$ are generators of SU$({\cal N})$, 
$G^a_{\mu\nu}=\partial_\mu A^a_{\nu}-\partial_\nu A^a_{\mu}+if^{a}_{~bs} 
A^b_{\mu} A^c_{\nu}$ is the field strength, $f_{abc}$ are the structure 
constants of SU$({\cal N})$ thereafter called the color gauge group. 

Quarks, labelled with $I$, possess color charges 
$Q_I$ in the adjoint representation of SU$({\cal N})$, $Q_I=Q^a_I\,T_a$.
These quantities can be expressed in 
terms of the basic variables $\eta_{Ij}$ in the fundamental representation,
\begin{equation}
Q_I=\sum_{a=1}^{{\cal N}^2-1}\sum_{i,j=1}^{{\cal N}}\, q_I^a\,\eta_{Ii}^{\ast}\left(T_a\right)^{i}_{\hskip0.5mm j}
\eta_{I}^{\hskip0.3mm j} 
\,.
\label
{Q-in-terms-eta}
\end{equation}                                           

The Euler--Lagrange equations for $\eta$ and $\eta^{\ast}$ read
\begin{eqnarray}
{\dot\eta}^i=-({\dot z}\cdot A^a)\left(T_a\right)^i_{\hskip0.5mm j}\eta^j\,\,,
\nonumber\\ 
{\dot\eta}^{\ast}_j=\eta^{\ast}_i\,({\dot z}\cdot A^a)\left(T_a\right)^i_{\hskip0.5mm j}\,\,.
\label
{Euler-Lagrange-eta-YMW}
\end{eqnarray}                                          
They can be combined into the {Wong equation} for the color charge 
evolution \cite{Wong},
\begin{equation}
{\dot Q}_a=-if_{abc}\,Q^b\left({\dot z}\cdot A^c\right).
\label
{8}
\end{equation}                                           
It is convenient to rescale the color variables: $Q\to-ig\,Q^a T_a$, $A_\mu\to(i/g)A_\mu^a T_a$, 
where $g$ is the Yang--Mills coupling constant.
Then Eq.~(\ref{8}) becomes 
\begin{equation}
\dot Q=ig\left[Q,\,{\dot z}^\mu A_\mu \right].
\label
{8--}
\end{equation}                                           
It follows that the color charge $Q$ shares with a top the property of precessing 
around some axis in the color space.

Varying $z^\mu$ and $A^\mu$ in the action (\ref{Balachandr-action}) gives the 
dynamical equations for respectively the Yang--Mills field and quarks:
\begin{equation}
D^\lambda G_{\lambda\mu}=4\pi g\sum_{I=1}^K\,\int_{-\infty}^\infty d\tau_I\,Q_I(\tau_I)\,
{\dot z}^I_\mu(\tau_I)\,\delta^4\left[x-z_I(\tau_I)\right],
\label
{YM-eq-}
\end{equation}                                          
\begin{equation}
m_0^I\,{\ddot z}_I^\lambda =
 {\dot z}^I_\mu\,{\rm tr}\left[Q_I\, G^{\lambda\mu}(z_I)\right].
\label
{Euler-Lagrange-z-YMW-}
\end{equation}                                          

In contrast to the electric charge $e$, which is a constant, the color charge 
$Q$ is a dynamical variable governed by the Wong equation (\ref{8--}).
Note, however, that the color charge magnitude is a constant of motion,
\begin{equation}
\frac{d}{ds}|{Q}|^2=2{\dot Q}^aQ_a=0\,,
\label
{Q-square}
\end{equation}                                          
which can be readily seen from (\ref{8}) written in the Cartan basis in which 
$f_{abc}=-f_{bac}$.
Furthermore, there is good reason to look for solutions of 
the Yang--Mills equations (\ref{YM-eq-}) satisfying the condition
\begin{equation}
Q^a(s)= {\rm const}.
\label
{Q-a=const}
\end{equation}                                          
Abandoning this condition would pose the problem of an infinitely rapid 
precession of $Q$ in view of the fact that the retarded field $A_\mu$ is 
singular on the world line.

It is clear from Eqs.~(\ref{YM-eq-}) and (\ref{Euler-Lagrange-z-YMW-}) that 
the action--reaction principle holds in the Yang--Mills--Wong theory.
If one conceives that only a single quark is in the universe, then 
$Q$ measures both the variation of the quark state for a given 
field state and variation of the field state for a given quark state.

Again, we look at self-interaction for revealing the relation between the 
action--reaction principle and energy-momentum conservation in an 
explicit form.
The strategy here copies that in the Maxwell--Lorentz electrodynamics, but has 
several traits associated with the fact that the field equations (\ref{YM-eq-}) 
are nonlinear.

There are {\it two kinds of retarded solutions} to the 
Yang--Mills equations, Abelian and non-Abelian \cite{k98}.
We first turn to the simplest case that the SU$(2)$ Yang--Mills field is 
generated by a single quark moving along an arbitrary timelike smooth world 
line \cite{k91}.
The retarded Abelian solution
\begin{equation}
A^\mu=q T_3\,{{\dot z}^\mu\over\rho}\,
\label
{Abelian-1-Pauli-}
\end{equation}                                       
resembles the Li\'enard--Wiechert solution of the Maxwell--Lorentz 
electrodynamics, whereas the retarded non-Abelian solution is given by
\begin{equation}
A^\mu=\mp\,{2i\over g}\,T_3\,{{\dot z}^\mu\over\rho}+
i\kappa\left(T_1\pm iT_2\right)R^\mu\,.
\label
{Nabelian-1-Pauli-}
\end{equation}                                       
Here $T_a$, $(a=1,2,3)$, are the generators of SU$(2)$, and $q$ and $\kappa$
are arbitrary real nonzero parameters.

A remarkable feature of retarded non-Abelian solutions bearing on our discussion 
is that the Yang--Mills equations determine not only the field, but also the 
color charge that generates this field, as exemplified by (\ref{Nabelian-1-Pauli-}).
This solution admits only a {\it single} value for the magnitude of the color 
charge carried by the quark \cite{k91}, \cite{k92},
\begin{equation}
|{Q}|^2=-\frac{4}{g^2}\,.
\label
{Q-square=2-g}
\end{equation}                                          
Recall that the electric charge $e$ of any particle in the 
Maxwell--Lorentz electrodynamics may be arbitrary.
The selection of a special magnitude for the color charge of the source takes
place also for a closed system of $K$ quarks evolving in the non-Abelian regime
\cite{k98},  
\begin{equation} 
{\rm tr}\,(Q^2_I)=- {4\over g^2}
\left(1-\frac{1}{{\cal N}}\right),
\quad
{\cal N}\ge 3\,.
\label
{Q-2=4/g-2(1-1/N)}
\end{equation}                       
Clearly this feature of the non-Abelian dynamics offers no danger to the 
fulfilment of the action--reaction principle.

In the {\it Abelian regime}, the field equations (\ref{YM-eq-}) 
linearize\footnote{The reason for this linearization is that the color 
field variables are restricted to the Cartan subalgebra of the Lie algebra of 
the associated gauge group.
Since the Lie algebra su$({\cal N})$ is of rank ${\cal N}-1$, there exist  ${\cal N}-1$
diagonal matrices $H_a$ forming a Cartan subalgebra of commuting
matrices.}, and hence, their retarded solution shows up as that in 
(\ref{F-mu-nu-LW-III})--(\ref{V-}).
All results of the previous section are reproduced with the only 
replacement $e^2\to q^2$.
The degrees of freedom appearing in the action (\ref{Balachandr-action}) are rearranged on 
the extremals subject to the retarded condition to give a dressed quark and 
Yang--Mills radiation, closely  resembling such entities in electrodynamics.
The behavior of a dressed quark is governed by the 
Lorentz--Dirac equation (\ref{LD}), which can be converted to the local 
energy-momentum balance (\ref{balance-LD}).

In the {\it non-Abelian regime}, the field equations (\ref{YM-eq-}) remain nonlinear, 
and superposing their solutions ceases to be true.
Aside from the one-quark solution (\ref{Nabelian-1-Pauli-}), there is need
to examine $K$-quark solutions, $K\ge 2$.
A consistent Yang--Mills--Wong theory can be formulated for the color gauge group 
SU$({\cal N})$ with ${\cal N}\ge K+1$ \cite{k98}. 
As an illustration we refer to a retarded SU$(3)$ field due to two
quarks \cite{k93}, \cite{k94}, 
\begin{equation}
A^\mu =\mp{2i\over g}\left(H_1\,{{\dot z}^\mu_1\over\rho_1}+
g\,\kappa\,E_{13}^\pm\,R_1^\mu\right)\,
\mp\,{2i\over g}\left(H_2\,{{\dot z}^\mu_2\over\rho_2}+g\,\kappa\,E_{23}^\pm\,
R_2^\mu \right).
\label
{alternative-sol-}
\end{equation}                                       
Here, $H_a$ and $E_{ab}$ are generators of SU$(3)$ in the Cartan--Weyl basis,
which are expressed in terms of the Gell-Mann matrices as follows: 
\begin{equation}
H_1= \frac12\left(\lambda_3+\frac{\lambda_8}{\sqrt{3}}\right),
\,\,
H_2= -\frac12\left(\lambda_3-\frac{\lambda_8}{\sqrt{3}}\right),
\,\,
E_{13}= \frac12\left(\lambda_4+i\lambda_5\right),
\,\,
E_{13}= \frac12\left(\lambda_6+i\lambda_7\right).
\label
{H-E-in-terms-lambdas}
\end{equation}                                       
$R_1^\mu=x^\mu-{z}^\mu_1(\tau_1)$ and $R_2^\mu=x^\mu-{z}^\mu_2(\tau_2)$ 
are, respectively, the four-vectors drawn from points ${z}^\mu_1(\tau_1)$ and 
${z}^\mu_2(\tau_2)$ on the world lines of quarks 1 and 2, where the signals 
were emitted, to the point $x^\mu$, where the signals were received.

Observing that $A^\mu$ is the {\it sum} of two single-quark terms, one may wonder 
of how the nonlinearity of the Yang--Mills equations is compatible with this 
fact.
The answer is simple: two  single-quark vector potentials with the {\it fixed}
magnitudes of the color charges, as shown in Eq.~(\ref{Q-2=4/g-2(1-1/N)}), are 
combined in Eq.~(\ref{alternative-sol-}), but it is impossible to build 
solution as an {arbitrary} superposition of these terms.
If either of them is multiplied by a coefficient different from 1 
and added to another, no further solution arises. 

Due to this feature -- which is characteristic of the general
$K$-quark case -- we have
\begin{equation}
\Theta^{\mu\nu}=\sum_{I}\left( \Theta_{~I}^{\mu\nu}+
 \sum_{J\ne I}\Theta_{~IJ}^{\mu\nu}\right), 
\label
{Theta-self-mix-decomp-}
\end{equation}                        
where $\Theta_{I}^{\mu\nu}$ is comprised of the field generated by
the $I$th quark, and $\Theta_{~IJ}^{\mu\nu}$ contains
mixed contributions of the fields due to the $I$th and $J$th quarks.
Furthermore,   $\Theta_{I}^{\mu\nu}$ splits into bound  and radiated parts.
Every term of Eq.~(\ref{Theta-self-mix-decomp-}) satisfies the local conservation 
law of the type shown in Eq.~(\ref{cons-Theta-I-II}), and hence represents a dynamically 
independent part of  $\Theta^{\mu\nu}$.
The stress-energy tensor $\Theta^{\mu\nu}$ is thus {\it similar in structure} 
to that in the Maxwell--Lorentz theory.

We now restrict our attention to a {\it single} quark of this $K$-quark system. 
For notational convenience, we omit the quark labelling.

Using the line of reasoning developed in the previous section, and
observing that the linearly rising term of $A_\mu$ does not contribute to 
$\Theta^{\mu\nu}$ because
\begin{equation}
{\rm tr}\,(H_l\,E^\pm_{mn})=0\,,
\quad
{\rm tr}\left(E^\pm_{kl}\,E^\pm_{mn}\right)=0\,,
\label
{tr-}
\end{equation}                                    
we arrive at the conclusion that the four-momentum of the
retarded Yang--Mills field generated by the quark under study is given by 
$P^\mu=P_{\rm bound}^\mu+{\cal P}^{\mu}$, where
$P_{\rm bound}^\mu$ and ${\cal P}^{\mu}$ are respectively the bound 
and radiated parts of this four-momentum.

An accelerated quark emits  
\begin{equation}
{\cal P}^{\mu}= -\frac23\,{\rm tr}\,(Q^2)\int^{{s}}_{-\infty}d\tau\,
{\ddot z}^2 \,{\dot z}^\mu\,.
\label
{P-rad-quark-non-A}
\end{equation}                        
Owing to the negative norm of the color charges, Eqs.~(\ref{Q-square=2-g}) and 
(\ref{Q-2=4/g-2(1-1/N)}), the emitted energy is negative, which suggests that
the quark gains, rather than loses, energy by emitting the Yang--Mills radiation 
in the non-Abelian regime.
An explicit calculation shows that this is indeed the case:
\begin{equation}
{\dot{\cal P}}\cdot {\dot z}= \frac{8}{3g^2}\left(1-\frac{1}{{\cal N}}\right)
{\ddot z}^2<0\,. 
\label
{P-rad-quark-cold}
\end{equation}                        
This phenomenon might be interpreted as absorbing convergent waves of positive 
energy rather than emitting divergent waves of negative energy \cite{k92}.

Adding the bound part of the field four-momentum
\begin{equation}
P^{\mu}_{{\rm bound}}={\rm tr}\,(Q^2)\left({1\over 2\epsilon}\,{\dot z}^{\mu} 
-
{2\over 3}\,{\ddot z}^{\mu}\right)
\label
{P-bound-c}
\end{equation}                        
to the mechanical four-momentum  $p^{\mu}_{0}=m_0{\dot z}^{\mu}$,
and carrying out the renormalization of mass in a way similar to (\ref{m-ren}),
gives the four-momentum of a dressed quark
\begin{equation}
p^\mu=m\left({\dot z}^\mu+\ell\, {\ddot z}^\mu\right).
\label
{p-mu-dress-quark}
\end{equation}                                          
Here,  $m$ is the renormalized mass, and
\begin{equation}
\ell
=
\frac{8}{3m g^2}
\left(1-\frac{1}{{\cal N}}\right)
\label
{tau-0-quark}
\end{equation}
is a characteristic length inherent in the non-Abelian dynamics of this dressed quark.
  
The mixed terms in Eq.~(\ref{Theta-self-mix-decomp-}) have integrable singularities 
$\rho^{-2}$ on every world line.
Their treatment is  therefore similar to that of $\Theta_{\rm mix}^{\mu\nu}$ in
the Maxwell--Lorentz electrodynamics.
The integration of these terms gives 
\begin{equation}
{\wp}^{\mu}=
-\int^{{s}}_{-\infty}d\tau\,f^{\mu}_{\rm ext}\!\left[{z}(\tau)\right],
\label
{P-mix-Wong}
\end{equation}                        
where the integrand is the {color} four-force exerted on the given quark by all other quarks at the instant $\tau$.
The explicit form of  $f^\mu_{\rm ext}$ is of no concern in the present context.

We reiterate mutatis mutandis the argument of the previous section to find
\begin{equation}
{\dot p}^\mu+{\dot{\cal P}}^{\mu}=f^{\mu}_{\rm ext}\,.
\label
{local-balance-YMW}
\end{equation}                        
According to this balance equation, the four-momentum 
$d{\wp}^{\mu}=-f^\mu_{\rm ext} ds$ extracted from an external field is used for
changing the four-momentum of the dressed quark $dp^{\mu}$ and emitting the 
Yang--Mills radiation  four-momentum ${d{{\cal P}}}^\mu$.
A special feature of equation (\ref{local-balance-YMW}) is that ${d{{\cal P}}}^0$ is 
associated with emitting negative-energy waves or, what is the same -- absorbing 
positive-energy waves.

Substitution of (\ref{p-mu-dress-quark}), (\ref{P-rad-quark-non-A}), and 
(\ref{P-mix-Wong}) into (\ref{local-balance-YMW}) gives the
equation of motion for a dressed quark 
\begin{equation} 
m\left[ {\ddot z}^\mu+\ell\left({\stackrel{\ldots}z}^{\hskip0.3mm\mu}+{\dot z}^\mu {\ddot z}^2\right)\right]=
f^{\mu}_{\rm ext}\,,
\label
{LD-quark-cold}
\end{equation}                       
differing from the Lorentz--Dirac equation (\ref{LD})
only in the overall sign of the parenthesized term and changing $\tau_0$ by 
$\ell$. 
If $f^\mu_{\rm ext}=0$,  the general solution of equation (\ref{LD-quark-cold}) 
is
\begin{equation} 
{\dot z}^\mu(s)=V^\mu\cosh(\alpha_0+w_0\ell\, e^{-s/\ell})
+U^\mu\sinh(\alpha_0+w_0\ell\,e^{-s/\ell})\,,
\label
{self-deceleration}
\end{equation}                         
where $V^\mu$ and $U^\mu$ are constant four-vectors such that 
$V\cdot U=0$,  $V^2=-U^2=1$, 
and $\alpha_0$ and $w_0$ are arbitrary parameters.
A free quark may therefore execute a non-uniform motion with 
exponentially decreasing {acceleration}.
The world line of this self-decelerated motion asymptotically approaches a 
straight line.
This situation can be interpreted in the spirit of the action--reaction
principle. 
Equation (\ref{local-balance-YMW}) becomes
\begin{equation}
d{p}^\mu=-d{\cal P}^{\mu}\,.
\label
{local-balance-single-quark}
\end{equation}                        
The free dressed quark feels the `reverse' four-momentum transfer 
(responsible for the self-deceleration) because the phenomenon of radiating the 
Yang--Mills four-momentum is actually changed by that of absorbing this 
four-momentum.
Nevertheless, Eq.~(\ref{local-balance-single-quark}) offers direct evidence 
that the action--reaction principle is equivalent to energy-momentum 
conservation on the world line.
 
\section{Gravitation}  
\label
{Gravitation}
The action--reaction  principle does not hold in the gravitational interaction
described by general relativity. 
Indeed, the coupling between a particle of mass $m$ and the gravitational field 
is equal to $m$, so that the particle is governed by the geodesic equation 
\begin{equation}
\frac{d^2z^\lambda}{d\tau^2}+\Gamma^\lambda_{~\mu\nu}\frac{dz^\mu}{d\tau}
\frac{dz^\nu}{d\tau}=0\,\,,
\label
{geodesics}
\end{equation}
which is mass-independent.
On the other hand, the field equation 
\begin{equation}     
R^{\mu\nu}-\frac12Rg^{\mu\nu}-8\pi G_{\rm N} t^{\mu\nu}=0\,\,
\label
{Hilbert-Einstein} 
\end{equation}
with the delta-function source 
\begin{equation}
t^{\mu\nu}(x)=m\int_{-\infty}^\infty d\tau\,{\dot z}^\mu(\tau)\,{\dot z}^\nu(\tau)\,
\delta^{4}[x-z(\tau)]\,\,
\label
{ideal gas}
\end{equation}
shows that the greater is $m$, the stronger is the generated gravitational field.
The influence of particles on the state of the gravitational field is different for 
different $m$, even though the gravitational field exerts on every particle in 
a uniform way, no matter what is  $m$.
This is contrary to the action--reaction principle.

Does this violation of the action--reaction principle imply that the 
energy--momentum conservation law is missing from general relativity?
While on the subject of arbitrary curved manifolds, the idea of translational 
invariance is irrelevant, whence it follows that not only energy and momentum 
are not conserved, but also the very construction
of energy and momentum suggested by Noether's first theorem is no longer  
defined.

To avoid this conclusion, one normally turns to field-theoretic treatments of 
gravity. 
This is feasible if the gravitational field can be granted to be `sufficiently 
weak',
\begin{equation}
g_{\mu\nu}=\eta_{\mu\nu}+\phi_{\mu\nu}\,, 
\label
{weak gravity}
\end{equation}
where $ |\phi_{\mu\nu}|\ll 1$.
The quantity $\phi_{\mu\nu}$ is thought of as a second-rank tensor 
field\footnote{Recent developments in bimetric theories of gravitation 
is reviewed in \cite{Schmidt-May}.} defined in 
a flat background ${\mathbb R}_{1,3}$ whose symmetry properties enable us to 
endow the resulting dynamics with conserved energy-momentum through the standard 
Noether's prescription.  

It is believed that general relativity leaves room for both weak and strong gravity. 
Strange though it may seem, a simple and convincing 
{criterion} for discriminating between {\it weak} and {\it strong} gravity
still remains to be established.
We therefore have to address this issue.
But our concern here is not with elaborating this criterion in every respect.
Rather, we only state the central idea and explicate it by the example of the 
Schwarzschild metric.

Intuition  suggests that the strong gravity should be associated with a great 
warping of spacetime.
However, a characteristic curvature whereby the changes in spacetime 
configurations might be rated as `drastic' is absent from general relativity, 
sending us in search of another measure of such changes. 
It seems reasonable to assume that switching between weak and strong gravitational
regimes is due to spacetime {topology} alterations.  
The `strong gravitational field' is then recognized as a qualitative rather 
than quantitative concept.
The field equations (\ref{Hilbert-Einstein}), being differential equations, are 
local in character.
They tell nothing about the topology of their solutions.
A global solution can be recovered when its infinitesimal pseudoeuclidean 
fragments are assembled into an integral dynamical picture, and the 
topology of this solution may well differ from the topology of Minkowski 
spacetime if the assembly is subject to a restrictive  boundary condition.
To illustrate, we refer to the Schwarzschild metric \cite{Schwarzschild},
\begin{equation}
ds^2=-\left(1-\frac{r_{\rm S}}{r}\right)dt^2+ 
\left(1-\frac{r_{\rm S}}{r}\right)^{-1}dr^2+
r^2d{\Omega}\,,
\quad
\label
{Schwarzschild-metric}
\end{equation}                                          
where, $d{\Omega}$ is the round metric in a sphere $S^2$, and 
$r_{\rm S}={2G_{\rm N}M}$ is the Schwarzschild radius. 
A 3-dimensional spacelike surface $\Sigma_3$ endowed with this metric has 
a twofold geometric interpretation.
First, it looks like a `bridge' between two otherwise Euclidean spaces, and, 
second, it may be regarded as the `throat of a wormhole' 
connecting two distant regions in one Euclidean space in the limit when this 
separation of the wormhole mouths is very large compared to the circumference 
of the throat  \cite{FullerWheeler}.

To describe a curved manifold ${\cal M}$, a set of overlapping coordinate 
patches covering  ${\cal M}$ is called for.
If one yet attempts to use a single coordinate patch, a singularity in the 
resulting description can arise.
The gravitation is amenable to a field-theoretic treatment until the mapping of 
the metric $g_{\mu\nu}$ into the field $\phi_{\mu\nu}$, as shown in (\ref{weak gravity}), 
is bijective and smooth, which is the same as saying that every curved spacetime 
configuration, associated with some gravitational effect, can be 
smoothly covered with a single coordinate patch.  
In contrast, for a manifold whose topology is nontrivial, the quest for a 
single-patch covering culminates in a singular boundary, bearing some resemblance to 
a shock wave, as exemplified by the Schwarzschild metric (\ref{Schwarzschild-metric}) 
in which the coefficient of $dr^2$ becomes singular 
at $r=r_{\rm S}$, so that this solution exhibits a standing spherical shock wave
of the gravitational field. 

One may argue that this is an apparent singularity, related to the choice of 
coordinates, because the curvature invariants are finite and well behaved at 
$r=r_{\rm S}$, and, furthermore, the equation for the geodesics 
(\ref{geodesics}) shows a singular behavior only at $r=0$.
There are coordinates $u$ and $v$, proposed in \cite{Fronsdal} and \cite{Kruskal}, 
\begin{equation}
u=\sqrt{\frac{r}{r_{\rm S}}-1}\exp\left(\frac{r}{r_{\rm S}}\right)\cosh\left(\frac{t}{r_{\rm S}}\right),
\quad
v=\sqrt{\frac{r}{r_{\rm S}}-1}\exp\left(\frac{r}{r_{\rm S}}\right)\sinh\left(\frac{t}{r_{\rm S}}\right),
\quad
\label
{Kruskal}
\end{equation}                                          
such that the Schwarzschild metric, being written in terms of  $u$ and $v$, is 
regular in the whole  $(u,v)$ plane, except for the point $v^2-u^2=1$ 
corresponding to $r=0$. 

In response to this objection, we would note that the introduction of these $u$ 
and $v$ is a clever trick to drive the shock wave in the singular point $r=0$.  
However, our prime interest is with the very existence of a shock wave, as 
evidence of the nontrivial topology, rather than its position in a particular 
coordinate system. 
The apparent regularity of the metric everywhere except $r=0$ is due to an 
unfortunate choice of coordinates which hides the Schwarzschild shock wave. 
 
The `floating' position of the shock wave makes it clear that the strong 
gravitational regime is unrelated to the magnitude of field variables.
It is a topologically nontrivial affair which renders the regime strong.

This brings up the question as to whether the violation of regular behavior of the 
Schwarzschild metric at $r=r_{\rm S}$ is an artefact of the original 
Schwarzschild description.
Some fifty years ago people were inclined to believe that such is indeed the case. 
By now, however, $r_{\rm S}$ is recognized as an objectively existing entity to
characterize the event horizon of an isolated spherically symmetric stationary 
black hole. 
The event horizon of a Schwarzschild black hole shows a demarcation between 
spacetime regions characterized by opposite signatures of the metric\footnote{
Note that the only quantity which has
a discontinuity jump at the front of a strong gravitational shock wave is the 
signature because the metric immediately anterior and posterior to the front 
can be brought to either of two diagonal forms: ${\rm diag}(+---)$ or ${\rm diag}(-+++)$.}.
{This geometrical layout, if it exists, provides an explicit scheme for 
interfacing the classical and the quantum} \cite{k8}.

Let us take a closer look at why energy and momentum are to be regarded as 
poorly defined concepts.
General relativity allows the Hamiltonian formulation for
at least such systems whose geometric rendition is compatible with the idea of 
asymptotically flat spacetime \cite{ADM-1}--\cite{ADM-3}.
More specifically, one supposes that the 
metric $g_{\mu\nu}$ approaches the Lorentz metric $\eta_{\mu\nu}$  at spatial 
infinity sufficiently rapidly, namely
\begin{equation}
g_{\mu\nu}=\eta_{\mu\nu}+O\left(\frac{1}{r}\right)\,,
\quad
\partial_{\lambda}g_{\mu\nu}=O\left(\frac{1}{r^2}\right)\,,
\quad
r\to\infty.
\label
{g-eta=1/r}
\end{equation}                                          
The second condition is claimed to be needed so that the Lagrangian 
\begin{equation}
L=\int d^3{\bf x}\,{\cal L}(t,{\bf x})\,
\label
{Lagr-}
\end{equation}                                          
with the commonly used first order Lagrangian density of the gravitational 
field sector
\begin{equation}
{\cal L}=\sqrt{-g}\,g^{\mu\nu}\left(\Gamma_{~\mu\nu}^{\sigma}\, 
\Gamma_{~\sigma\lambda}^{\lambda}
-\Gamma_{~\mu\sigma}^{\lambda}\, \Gamma_{~\nu\lambda}^{\sigma}\right)\,
\label
{Lagr-function}
\end{equation}                                          
should converge.
The volume integral in (\ref{Lagr-}) diverges for the 
Schwarzschild solution expressed in terms of the original Schwarzschild 
coordinates, appearing in (\ref{Schwarzschild-metric}), because ${\cal L}=O(1)$ 
as $r\to\infty$.
In contrast, the use of isotropic coordinates, which recasts the Schwarzschild
metric (\ref{Schwarzschild-metric}) into   
\begin{equation}
ds^2=-\frac{1-\frac{r_{\rm S}}{4r}}{1+\frac{r_{\rm S}}{4r}}\,dt^2+
\left(1+\frac{r_{\rm S}}{4r}\right)\left(dx^2+dy^2+dz^2\right),
\quad
\label
{Schwarzschild-isotropic}
\end{equation} 
results in ${\cal L}=O(1/r^4)$, and hence affords the convergence of the 
Lagrangian (\ref{Lagr-}). 
It follows from this simple example that both the asymptotical flatness  
\begin{equation}
R^\alpha_{~\beta\gamma\delta}\to 0\,, 
\quad 
r\to \infty\,,
\label
{flat}
\end{equation}                      
and a good choice of coordinates share in the responsibility for the convergence of additive quantities such as the
Lagrangian and Hamiltonian.                                          

With the neat Hamiltonian formulation developed in \cite{ADM-1}--\cite{ADM-3}, 
one would expect that the total energy-momentum $P^\mu$ has 
an unambiguous significance. 
We now examine the correctness of this expectation restricting ourselves to $P^0$ for simplicity.

The total energy is given by the numerical value of the Hamiltonian 
\begin{equation}
E=\int d^3{\bf x}\,{\cal H}\left(t,{\bf x}\right).
\label
{Mass-}
\end{equation}                                          
${\cal H}$ is a cumbersome construction which is immaterial for our discussion.
However, the key part of this construction proves to be cast \cite{Regge} in a 
convenient form, 
\begin{equation}
E=\frac{1}{16\pi}\oint dS_j\left(\frac{\partial}{\partial x_i}\,g_{ij}-
\frac{\partial}{\partial x_j}\,g_{ii}\right).
\label
{Total-energy}
\end{equation}                                          
Here, the integral is evaluated over a 2-dimensional surface at spatial infinity.

It is possible to prove \cite{SchoenYau, Witten} that an isolated gravitating 
system having non-negative local mass density has non-negative total energy $E$.
For example,  for the Schwarzschild configuration generated by a point particle 
of mass $m$ the surface integral (\ref{Total-energy}) is easily evaluated to give
\begin{equation}
E=m.
\label
{Mass-Schwarzschild}
\end{equation}                                          

Could the condition (\ref{g-eta=1/r}) be relaxed so that the asymptotical 
flatness condition (\ref{flat}) would hold, and every pertinent additive 
quantity in this Hamiltonian formulation remains convergent?
To be more precise, we proceed from the metric $g_{\mu\nu}$ exhibiting the 
asymptotic behavior (\ref{g-eta=1/r}), and transform the initial spatial 
coordinates $x^i$ into new ones ${\bar x}^i$,
\begin{equation}
x^i={\bar x}^i\left[1+ f({\bar r})\right],
\label
{diffeomorphism-gen}
\end{equation}                                          
where $f$ is an arbitrary regular function subject to the following conditions:
\begin{equation}
f({\bar r})\ge 0\,,
\quad
\lim_{{\bar r}\to\infty} f({\bar r}) =0\,,
\quad
\lim_{{\bar r}\to\infty} {\bar r}\,f'({\bar r}) =0\,.
\label
{conditions_on_f}
\end{equation}                                          
For the mapping (\ref{diffeomorphism-gen}) to be bijective,
the condition
\begin{equation}
\frac{\partial r}{\partial{\bar r}}=1+f({\bar r})+{\bar r}f'({\bar r})>0\,
\quad
\label
{reversibl}
\end{equation}                                          
is necessary and sufficient.
Indeed, with (\ref{reversibl}), the mapping is explicitly invertible, 
\begin{equation}
J=\det \left(\frac{\partial x}{\partial{\bar x}}\right)=
\left[1+f({\bar r})\right]\frac{\partial r}{\partial{\bar r}}\ne 0\,.
\quad
\label
{Jacobian_ne_0}
\end{equation}                                          
One such example \cite{Denisov} is
\begin{equation}
f({\bar r})=2\,\alpha^2\,\sqrt{\frac{l}{{\bar r}}} 
\left[1-\exp\left(-\frac{\epsilon^2\,{\bar r}}{l}\right)\right],
\label
{diffeomorph}
\end{equation} 
where $\alpha$ and $\epsilon$ are arbitrary nonzero numbers, and $l$ is an 
arbitrary parameter of dimension of length.
This is a bijective monotonically increasing regular mapping $r\to{\bar r}$ 
which becomes $1$ as $\epsilon\to 0$.
The leading asymptotical terms of spatial components of the metric and those of 
the Christoffel symbols can be shown \cite{Denisov} to be 
\begin{equation} 
g_{ij}=\delta_{ij}+O\left(\frac{1}{{\bar r}^{1/2}}\right)\,,
\quad 
\Gamma_{~jk}^{i}=O\left(\frac{1}{{\bar r}^{3/2}}\right)\,,
\quad {\bar r}\to\infty\,,
\label
{g-eta=1/sqrt_r}
\end{equation} 
while the Lagrangian density behaves as
\begin{equation}
{\cal L}=O\left(\frac{1}{{\bar r}^{7/2}}\right),
\quad 
{\bar r}\to \infty\,.
\label
{Hamiltonian-transf-behavior}
\end{equation}                                          
This provides the convergence of the volume integral in (\ref{Lagr-}).
Other additive quantities prove to be convergent as well\footnote{Note also 
that the asymptotical flatness, Eq.~(\ref{flat}), is still the case.}. 

The mapping (\ref{diffeomorphism-gen}) with $f$ defined in (\ref{diffeomorph}) 
is instructive to apply to the Schwarzschild metric which is initially written 
in terms of isotropic coordinates (\ref{Schwarzschild-isotropic}).
One can show \cite{Denisov} that the total energy of the Schwarzschild configuration 
generated by a point particle of mass $m$ takes any positive values, greater 
than, or equal to $m$, when $\alpha^2$ runs through ${\mathbb R}_{+}$,
\begin{equation}
E=m\left(1+\alpha^4\right).
\label
{Mass}
\end{equation}                                          

We thus see that the total energy of gravitational systems with nontrivial 
topological contents depends on the foliation of spacetime.
The Schwarzschild solution expressed in terms of coordinates 
for which the asymptotic condition is given in a relaxed form, 
Eq.~(\ref{g-eta=1/sqrt_r}), is a good case in point.
This solution rearranges the initial degrees of freedom appearing 
in the Lagrangian density (\ref{Lagr-function}) to yield a
coordinatization-dependent expression for the total energy functional (\ref{Total-energy}).
The same is true of the associated momentum. 

The situation closely parallels that in the paradoxial Banach--Tarski theorem 
which states \cite{Banach}: given a unit ball in three dimensions, there exists 
a decomposition of this ball into a finite disjoint subsets which can then be 
reassembled through continuous movements of the pieces, without running into one 
another and without changing their shape, to yield another ball of larger 
radius.
These situations  share a common trait in that both the Banach--Tarski
decomposition and the Schwarzschild black hole formation are due to topological 
rearrangements which are responsible for making the three-dimensional measures 
of the resulting geometrical layouts poorly defined. 
The measure appearing in the Banach--Tarski theorem is the ordinary volume of  
the balls (more precisely, Lebesgue measure), while the measure 
in the gravitational energy problem is that of the functional (\ref{Mass-}).
When turning to the surface integral for calculation of the total energy, 
Eq.~(\ref{Total-energy}), there arises the situation which may be likened to 
that of the Hausdorff paradox on enlarging spheres \cite{Hausdorff}.

The usual inference that the Banach--Tarski partitioning procedure has nothing
to do with physical reality because there is no material ball which is not made 
of atoms overlooks one important instance -- black holes.
Each isolated, stationary black hole is completely specified by three parameters: 
its mass $m$, angular momentum $J$, and electric charge $e$.
Whatever the content of a system which collapses under its own gravitational 
field, the exterior of the resulting black hole is described by a Kerr--Newman 
solution.
All initial geometric features of the collapsing system, except for those
peculiar to a perfect ball, which may possibly rotate and carry electric charge, 
disappear in the black hole state \cite{Heusler, Chrusciel}. 
Furthermore, the event horizon which is meant for personifying the black 
hole is devoid of the grain structure that was inherent in the collapsing system.

\section{Discussion and outlook}
\label
{Discussion}
In Sects. 2 and 3 we saw that a careful analysis of the self-interaction problem may
give an insight into the relation between the action--reaction principle and 
energy-momentum conservation provided that the rearrangement of degrees of 
freedom is taken into account. 
Umezawa \cite{umezawa} was the first to put the term {`rearrangement'} in 
circulation by the example of spontaneous symmetry breaking.
The mechanism for rearranging classical gauge fields was further studied 
in \cite{k92}, \cite{k98}, \cite{k07}. 
While a precise formulation of this mechanism is still an open problem, the
intuitive idea underlying the rearrangement is quite simple.     
In choosing variables for the description of a field system, preference is 
normally given to those which are best suited for realizing all supposed 
fundamental symmetries of the action. 
But some degrees of freedom so introduced may be unstable.
This gives rise to reassembling the initial degrees of freedom into new, stable 
aggregates whose dynamics is invariant under broken or deformed groups of 
symmetries.
Aggregates obeying the usual requirement of stability 
\begin{equation}
\delta S=0\,,\quad \delta^2S>0\, 
\label
{stability}
\end{equation}                                          
form readily in field theories affected by spontaneous symmetry breaking, as 
exemplified by the Goldstone and Higgs models.
However, this criterion for discriminating between stable and unstable modes is 
difficult if not impossible to apply to local field theories with delta-function 
sources owing to divergences arising in the self-interaction problem.
We thus have to look for alternative criteria.

Let us return to the Maxwell--Lorentz electrodynamics.
We take, as the starting point, the on-shell dynamics of a bare particle and 
electromagnetic field engendered by the equations of motion ${\varepsilon}^\lambda=0$ 
and ${\cal E}_\mu=0$, Eqs.~(\ref{Eulerian-part}) and (\ref{Eulerian-em-II}), 
together with the retarded boundary condition.
However, this dynamics blows up on the world line, which, in view of 
Eq.~(\ref{Noether-1-id-em}), is tantamount to stating that the measure 
$T^{\mu\nu}d\sigma_\nu$ is ill defined. 
It would be tempting to construe such divergences as evidences of instability. 

We then divide $T^{\mu\nu}d\sigma_\nu$ into a well-defined part and the remainder.
But this separation is ambiguous: an arbitrary regular term can be added 
to one part and subtracted from the other to give an equivalent separation. 
To fix the separation, we impose the condition that every term obey the local 
conservation law (\ref{cons-Theta-I-II}).    
The functionals (\ref{P-rad-def}) and (\ref{P-mix}), expressing, respectively, 
the four-momentum radiated by the charge and four-momentum extracted from a 
{free} field, refer to the well-defined part of $T^{\mu\nu}d\sigma_\nu$.
We complete the definition of 
$\left(\Theta_{\rm bound}^{\mu\nu}+t^{\mu\nu}\right)d\sigma_\nu$ by carrying out 
the renormalization of mass, Eq.~(\ref{m-ren}).
The functional (\ref{p-dressed}) is regarded as the four-momentum of a dressed 
charged particle. 
As might be expected, the rearrangement outcome, the Lorentz--Dirac 
equation $\Lambda^\mu=0$, Eq.~(\ref{LD}), governing the behavior of the dressed 
particle, is depleted of some symmetries embedded in the action.
Indeed,  this dynamical equation is not invariant under time reversal 
$s\to -s$.

We thus come to a new on-shell dynamics in which the equation of motion for a
bare particle ${\varepsilon}^\mu=0$ is replaced with the equation of motion 
for a dressed particle $\Lambda^\mu=0$, and all relevant integral quantities 
are well defined.
Therefore, the rearrangement of the Maxwell--Lorentz electrodynamics can be briefly
outlined as follows: since the on-shell dynamics which arises from 
extremizing the action and imposing the retarded boundary condition is divergent, 
the initial degrees of freedom appearing in the action are induced to reassemble 
into new aggregates governed by a tractable dynamics.
What are the ways open to this reassembly?

The arena for rearranging the Maxwell--Lorentz electrodynamics is a line 
${\mathbb R}$ covered by the evolution variable $\tau$ which parametrizes the 
world line, and a plane ${\mathbb E}_2$ spanned by two vectors $R^\mu$ and 
$V^\mu$ used in defining the retarded Li\'enard--Wiechert 2-form 
$F$, Eq.~(\ref{F-mu-nu-LW-III}). 
Reparametrization invariance of the action and local SL$(2,{\mathbb R})$ 
invariance of the 2-form $F$ control the rearrangement scenario.
Hence, the ways open to the reassembly are specified by the properties of 
the local translation group $T$ responsible for reparametrizations and the 
local SL$(2,{\mathbb R})$ group acting in the retarded field plane.

Recall the main implication of reparametrization invariance, Noether's second 
theorem \cite{Noether}.
It is convenient to restrict our consideration to an infinitesimal 
reparametrization
\begin{equation}
\delta\tau=\epsilon (\tau)\,,
\label
{reparametrization-infinitesimal}
\end{equation}                        
where $\epsilon$ is an arbitrary smooth function of $\tau$ close to zero,
which becomes vanishing at the end points of integration.
Variation of $\tau$ implies the corresponding variation of the world line 
coordinates
\begin{equation}
\delta z^\mu={\dot z}^\mu\epsilon\,.
\label
{reparam-infsm-coord}
\end{equation}                        
In response to the reparametrization (\ref{reparametrization-infinitesimal})--(\ref{reparam-infsm-coord}),
the action varies as 
\begin{equation}
\delta S=
\int\! d\tau\,{\varepsilon}_\mu{\dot z}^\mu\epsilon\,.
\label
{variation-S-sec2.6}
\end{equation}                        
Let $S$ be invariant under reparametrizations, $\delta S=0$.
Because $\epsilon$ is assumed to be an arbitrary function $\tau$, one concludes 
that
\begin{equation}
{\dot z}^\mu{\varepsilon}_\mu=0\,.
\label
{linear-dependence-Eulerians}
\end{equation}                        
This equation is a manifestation of Noether's second theorem:  
invariance of the action under the transformation group (\ref{reparametrization-infinitesimal}) 
involving an arbitrary infinitesimal function $\epsilon$ implies a linear 
relation between Eulerians. 

The identity (\ref{linear-dependence-Eulerians}) suggests that ${\varepsilon}_\mu$ 
contains the projection operator on a hyperplane with normal ${\dot z}^\mu$, 
\begin{equation}
\stackrel{\scriptstyle {\dot z}}{\bot }_{\hskip0.5mm\mu \nu }\,=\,{\eta }_{\mu \nu
}-\frac{{\dot z}_{\mu }{\dot z}_{\nu }}{{\dot z}^{2}}\,,  
\label
{projector-def}
\end{equation}
annihilating identically any vector parallel to ${\dot z}^\mu$.  
Reparametrization invariance bears on the projection structure of 
the basic dynamical law for a bare particle which can be written\footnote
{The projector $\stackrel{\scriptstyle {{\dot z}}}{\bot}$ may arise in (\ref{Newton}) from a completely different 
origin, namely smooth embedding of Newtonian dynamics into sections of Minkowski 
space perpendicular to the world line \cite{k07}.} as
\begin{equation} 
\stackrel{\scriptstyle {\dot z}}{\bot}({\dot p}-f)=0\,,
\label
{Newton}
\end{equation}                         
where $p$ is the four-momentum of a bare particle, and $f$ an external 
four-force.

In view of the identities ${\dot z}^2=1,\,{\dot z}\cdot {\ddot z}=0,\,
{\dot z}\cdot{\stackrel{\ldots}z}=-{\ddot z}^2$, the Lorentz--Dirac equation 
(\ref{LD}) can be brought to the form of Eq.~(\ref{Newton}) in which $p$ 
is the four-momentum of a dressed particle, defined in (\ref{p-dressed}), and 
$f$ is again an external four-force.

The structure of (\ref{Newton}) makes it clear that a dressed particle 
experiences only an external force.
This equation contains no term through which the dressed particle interacts with 
itself.
The rearrangement eliminates self-interaction.
The rearranged dynamical picture contains only autonomous, foreign to each other entities.

It may be worth pointing out that both equation of motion for a bare particle 
${\varepsilon}^\mu=0$ and  equation of motion for a dressed particle $\Lambda^\mu=0$ 
are generally not invariant under reparametrizations.
Instead, this local symmetry leaves its imprint on the form of ${\varepsilon}^\mu$
and $\Lambda^\mu$ through the presence of the projector 
$\stackrel{\scriptstyle {{\dot z}}}{\bot }$.

Invariance under the SL$(2,{\mathbb R})$ group stems from the fact 
that the 2-form $F$ describing the retarded field of a {single} charge is 
proportional to ${R}\wedge{V}$, that is, $F$ is decomposable \cite{k92, k07}.
Given a decomposable 2-form $F$, the invariant ${\cal P}=\frac12 F_{\mu\nu}{}^\ast\!F^{\mu\nu}$ is identically zero.
As for the invariant ${\cal S}=\frac12 F_{\mu\nu}F^{\mu\nu}$, using 
(\ref{F-mu-nu-LW-III})--(\ref{V-}), we find  ${\cal S}=-e^2/\rho^4$.
Therefore, a single charge moving along an arbitrary timelike world line 
generates the retarded field $F^{\mu\nu}$ of {electric} type.
In other words, whatever the motion of the charge, there is a Lorentz frame of 
reference, special for each point $x^\mu$, such that only electric field
persists, more precisely, $\vert{\bf E}\vert=e/\rho^2$ and ${\bf B}=0$. 

Rewrite (\ref{F-mu-nu-LW-III}) as
\begin{equation}
F=\frac{e}{\rho^2}\,\varpi\,,
\quad
\varpi={R}\wedge{V}\,.
\label
{F-LW-III}
\end{equation}                          
A pictorial rendition of the bivector $\varpi$ is the parallelogram of the
vectors $R^\mu$ and $V^\mu$.
The area $A$ of the parallelogram is  
\begin{equation}
A=\sqrt{-V^2\left(\stackrel{\scriptstyle V}{\bot}\!R\right)^{\hskip0.3mm 2}}
=V\cdot R=1\,.
\label
{A=sqrt(-V-2-bot-V-c)}
\end{equation}                          
The bivector $\varpi$ is invariant under the special linear group 
of real unimodular $2\times 2$ matrices SL$(2,{\mathbb R})$ which rotate and 
deform the initial parallelogram, converting it to parallelograms of unit area 
belonging to the plane spanned by the vectors
$R^\mu$ and $V^\mu$.  
Therefore, $\varpi$ is independent of concrete directions and magnitudes of 
the constituent vectors $R^\mu$ and $V^\mu$.  
$\varpi$ depends only on the parallelogram's orientation.
The parallelogram can always be built from a timelike unit vector 
$e_0^\mu$ and a spacelike imaginary-unit vector $e_1^\mu$ perpendicular to
$e_0^\mu$, $\varpi={e}_0\wedge{e}_1$.
In fact, there are three different cases: 

\noindent
(i) $V^2>0$,  
\begin{equation}
e_0^{\mu}=\frac{V^{\mu}}{\sqrt{V^2}}\,,
\quad 
e_1^{\mu}=\sqrt{V^2}\left(-R^{\mu}+\frac{V^\mu}{V^2}\right),
\label
{1.}
\end{equation}                         
(ii) $V^2<0$, 
\begin{equation}
e_0^{\mu}=\sqrt{-V^2}\left(R^{\mu}-\frac{V^\mu}{V^2}\right),
\quad 
e_1^\mu=\frac{V^{\mu}}{\sqrt{-V^2}}\,,
\label
{2.}
\end{equation}                          
(iii) $V^2=0$,
\begin{equation}
e_0^{\mu}=\frac{1}{\sqrt{2}}\left(\rho V^{\mu}+\frac{R^{\mu}}{\rho}\right), 
\quad 
e_1^{\mu}=\frac{1}{\sqrt{2}}\left(\rho V^{\mu}-\frac{R^{\mu}}{\rho}\right).
\label
{3.}
\end{equation}                          
In the Lorentz frame with the time axis parallel to the vector $e^\mu_0$, 
all components of $F^{\mu\nu}$ are vanishing, except for $F^{\hskip0.2mm 01}$.
The formulas (\ref{1.})--(\ref{3.}) specify explicitly a frame 
in which the retarded electromagnetic field generated by a single
arbitrarily moving charge appears as a pure Coulomb field at each observation
point.
With a curved world line, this frame is noninertial.

The decomposable 2-form  $F$ is invariant under the SL$(2,{\mathbb R})$ 
transformations which can be carried out independently at any spacetime point.
Therefore, we are dealing with {local} invariance.
This invariance is not pertinent to electrodynamics as a whole, and hence 
gives rise to no Noether identities.
Rather, this is a property of the {retarded solution} $F_{\rm ret}^{\mu\nu}$, shown in
Eq.~(\ref{F-mu-nu-LW-III})--(\ref{V-})\footnote{The advanced field 
$F_{\rm adv}^{\mu\nu}$ can also be represented in a form similar 
to (\ref{F-mu-nu-LW-III})--(\ref{V-}), that is, the 2-form $F_{\rm adv}$ is decomposable  
whereas combinations $F_{\rm ret}+\alpha\,F_{\rm adv}$ are not.}.

Therefore, the {retarded solution} $F_{\rm ret}^{\mu\nu}$ is determined not only 
by the field as such but also by the frame of reference in which this quantity 
is measured.  
On the other hand, $\Theta^{\mu\nu}$ is not invariant under such SL$(2,{\mathbb R})$ 
transformations; $\Theta^{\mu\nu}$ carries information about both
the field and the Lorentz frame which is used to describe $F_{\rm ret}^{\mu\nu}$. 
Nevertheless, the functionals (\ref{P-rad-def}), (\ref{P-mix}), and 
(\ref{p-dressed}) are well defined and frame-independent.

The rearrangement in the Yang--Mills--Wong theory shows a general resemblance of 
that in the Maxwell--Lorentz electrodynamics.
The field strength generated by a single quark is also given by a 
decomposable 2-form $F$ in both Abelian and non-Abelian regimes.
The retarded Yang--Mills field $F$ is always invariant under the local group SL$(2,{\mathbb R})$.

A special feature of the Yang--Mills--Wong theory (as opposed to the 
Maxwell--Lorentz electrodynamics) is that non-Abelian regimes of evolution 
exhibit {\it spontaneously deformed} gauge symmetries \cite{k98, k07}. 
Without going into detail, we explicate this phenomenon by the simplest 
example.
Consider the solution (\ref{Nabelian-1-Pauli-}) which describes the retarded 
non-Abelian field generated by a single quark in the SU$(2)$ Yang--Mills--Wong 
theory.
By introducing an alternative matrix basis
\begin{equation}
{\cal T}_1= T_1\,,
\quad
{\cal T}_2=iT_2\,,
\quad
{\cal T}_3= T_3\,,
\label
{sl(2,R) basis}
\end{equation}                                       
we convert this solution to the form $A_\mu ={\cal A}_\mu^a\,{\cal T}_a$
where all coefficients ${\cal A}_\mu^a$ are imaginary. 
Elements of this basis obey the commutation relations of the sl$(2,{\mathbb R})$ 
Lie algebra. 
We thus see that the gauge group of the solution (\ref{Nabelian-1-Pauli-}) is 
actually SL$(2,{\mathbb R})$\footnote{This SL$(2,{\mathbb R})$ gauge group should
not be confused with the SL$(2,{\mathbb R})$ symmetry transformations which 
leave a decomposable 2-form $F$ unchanged.
Given the initial SU$({\cal N})$ gauge symmetry with ${\cal N}\ge 2$, 
the spontaneously deformed gauge symmetry is found to be embedded in the SL$({\cal N},{\mathbb R})$
group \cite{k98, k07}.  
}.
Where does this group of symmetry come from?
Its origin bears no relation to spontaneous symmetry breakdown:
SU$(2)$ and SL$(2,{\mathbb R})$ are the {compact and noncompact real forms} of 
the complex group SL$(2,{\mathbb C})$. 
Invariance of the action under SU$(2)$ automatically entails its invariance 
under the complexification of this group, SL$(2,{\mathbb C})$.
The emergence of a solution invariant under a real form of 
SL$(2,{\mathbb C})$ different from the initial 
SU$(2)$ is a rearrangement phenomenon specific to the Yang--Mills--Wong theory,
called {spontaneous symmetry deformation}.
The solutions (\ref{Nabelian-1-Pauli-}) and (\ref{Abelian-1-Pauli-}) are
different not only in their symmetry aspect, but also in physical 
manifestations, say, the former manifests itself as the Yang--Mills field of 
`magnetic' type while the latter is the Yang--Mills field of `electric' type.
Dressed quark, associated with these solutions, are governed 
by respectively equations of motion (\ref{LD-quark-cold}) and (\ref{LD}),
both being in agreement with the action--reaction principle.

The rearrangement of general relativity is vastly different from that of the
Maxwell--Lorentz electrodynamics and Yang--Mills--Wong theory. 
Indeed, the total stress-energy tensor $T^{\mu\nu}$ is identical to the left-hand side 
of Eq.~(\ref{Hilbert-Einstein}), that is, the on-shell $T^{\mu\nu}$ is zero.
It is therefore impossible to define a three-dimensional measure weighted with
$T^{\mu\nu}$. 
And yet, the on-shell dynamics exhibits a kind of blow-up: gravitational singularities.
This troublesome feature of the theory is found even if delta-function sources 
are substituted by continuously distributed matter obeying a reasonable energy 
condition, the local positive energy condition \cite{Penrose, Hawking, PenroseHawking}.
However, the responsibility for the rearrangement does not rest with the 
divergent dynamics.
Gravitational degrees of freedom are induced to reassemble into new topologically
nontrivial aggregates due to instabilities which owe their origin to the failure 
of the action-reaction principle.   
The ways open to the rearrangement of general relativity are 
specified by the properties of four-dimensional reparametrizations
\begin{equation}
x^\mu=F^\mu(x')\,,
\quad
{g}_{\mu\nu}(x)=
\frac{\partial{x'}^{\alpha}}{\partial x^\mu}\,
\frac{\partial{x'}^{\beta}}{\partial x^\nu}\,{g'}_{\alpha\beta}(x')\,,
\label
{diffeomorph-brane}
\end{equation}                        
where $F^\mu$ are arbitrary smooth functions.
These transformations form an infinite group, the group of 
{\it diffeomorphism} invariance implying invariance of the metric 
under the local Lorentz group and parallel transport group.

It would be interesting to inquire into why the functionals (\ref{Mass-}) and 
(\ref{Total-energy}) become coordinatization-dependent for systems having 
nontrivial topological contents in the light of the analyses which are lumped 
together as the `Banach--Tarski theorem'  \cite{Wagon}. 
Note that the very analogy between the Banach--Tarski decomposition and the 
rearrangement of gravitational degrees of freedom may seem in doubt because the 
former has to do with sets of {\it points} in Euclidean space ${\mathbb E}_3$, 
whereas the latter refers to the pseudo-Riemannian {\it metric} structure. 
But the resemblance of these procedures is ensured by the fact that the
study of the affair with ${\mathbb E}_3$ is actually transferred to exploring 
the properties of bijective mappings of sets in  ${\mathbb E}_3$, and the like 
is true for the rearrangement of gravitational degrees of freedom.   
A central idea of the Banach--Tarski analyses 
is that if a bounded set can be decomposed in a paradoxial way with respect to 
a group $G$, 
then $G$ contains {\it free} subgroups, in particular a ball in 
${\mathbb E}_3$ is SO$(3)$-paradoxial because the action of SO$(3)$ is that of 
a {free non-Abelian} isometry group \cite{Wagon}.
The development of this idea in relation to the action of the isometry group 
composed of the local SO$(1,3)$ group and parallel transport group, 
having free non-Abelian subgroups, may give a plausible explanation for the fact that the 
measure of integral quantities such as (\ref{Mass-}) and (\ref{Total-energy}) 
is to be poorly defined.

On the other hand, the Banach's theorem stating that no paradoxial 
decompositions exists in ${\mathbb R}$ and ${\mathbb E}_2$  \cite{Wagon} should 
be likened to the affair with the well-defined measures in the rearranged 
Maxwell--Lorentz electrodynamics and Yang--Mills--Wong theory.
The class of groups whose actions preserve finitely additive, isometry-invariant 
measures of the bounded sets on ${\mathbb R}$ and ${\mathbb E}_2$ are 
known to be {\it amenable} groups, specifically {\it solvable} groups,
which include Abelian groups. 
It is conceivable that the groups of reparametrizations and local SL$(2,{\mathbb R})$ 
transformations controlling the rearrangement of these theories have what
amounts to the desired properties of amenable groups.  

A natural question may now arise: What is the reason for the existence of 
scenarios in which gravitational 
degrees of freedom reassemble in a topologically nontrivial fashion, say, into a 
Schwarzschild black hole, so that the
asymptotic condition (\ref{g-eta=1/r}) is met, and the total energy
functional (\ref{Total-energy}) becomes a well defined, non-negative quantity
\cite{SchoenYau, Witten}?
The suggestion can be made that the diffeomorphisms controlling such scenarios 
are restricted to the groups deprived of free subgroups.

Does the action--reaction principle remain its validity for quantum field 
theories such as quantum electrodynamics?
Three obstacles apparently placed in incorporation of this principle into the
quantum context are as follows:

$\bullet$ By virtue of vacuum polarization, the charge of a bare particle 
is no longer constant, but rather a time-varying dynamical quantity whose
numerical value is determined by virtual pair  screening.
It is unlikely that this {fluctuating} quantity may be taken to be a measure of
both variation of the electron state for a given electromagnetic field
state and variation of the state of electromagnetic field for a given electron state.    

$\bullet$ Heisenberg's uncertainty principles is contrary to bringing a 
{contact} interaction into coincidence with {exact} values of the 
four-momenta appearing in the local four-momentum balance.
In the quantum realm, the  four-momentum balance is either nonlocal or fuzzy.

$\bullet$ The rearrangement of initial degrees of freedom in the quantum picture
occurs much differently than in the classical picture. 
The criterion of stability, Eq.~(\ref{stability}), is {alien} to the 
quantum regime of evolution because any world line passing through the chosen 
end points -- and not just the world line which renders the action extremal -- 
contributes to the Feynman path integral. 
Therefore, the instability is of little, if at all, significance for the quantum 
rearrangement. 

However, it would be very strange if the Nature does reject the quantum utility 
of the principle which is so useful at the classical level.

\end{document}